# A Study on Web Application Vulnerabilities to find an optimal Security Architecture

Dr. C. Amuthadevi, Sparsh Srivastava, Raghav Khatoria and Varun Sangwan
SRM Institute of Science and Technology, Chennai, Tamil Nadu
amuthadc@srmist.edu.in, ss5451@srmist.edu.in, rk7664@srmist.edu.in and vs5019@srmist.edu.in

**ABSTRACT -**

Over the past three decades, computers have managed to make their way into a majority of households. Due to this enormous transition, the surge in the internet's popularity was inevitable. Just like everything else, whatever has a pro also has a con, both are faces of the same coin. Nowadays, web security has become one of the biggest challenges to the corporate world. All web applications are prone to security vulnerabilities, it's the developers job to follow the latest norms in order to effectively reduce the threat posed by unauthorised users or programs. In this paper, we have tried to analyse the major problems faced in the real world and have tried to come up with effective solutions to combat all the aforementioned problems.

*Keywords - Web Applications, data security, security vulnerabilities.*

## 1. INTRODUCTION -

Software security is a very important part of web application development. Generally when any application is being created (web or of any sort), they follow a software development life cycle. Software security is generally overlooked during the process of development and has often shown to have a detrimental effect. The application's security should be enforced early in the software development life cycle to avoid the considerable expense and effort of adding security later.

Over the years developers have chosen the web as their top pick for creating and deploying applications due to its cross platform compatibility. But this hasn't been without its drawbacks. There are several security vulnerabilities that exist which can have a severe effect on the developed application. Some of the most notorious of the buch include, Injection flaws, Broken authentication, Sensitive data exposure, Broken access controls and Cross-site scripting (XSS). According to an article published by 'positive technologies' titled Web Applications vulnerabilities and threats: statistics for 2019, unauthorised access to applications is still possible on 39 percent of sites and breaches of sensitive data were a threat in 68 percent of web applications. Another very interesting stat from 'Tech beacon' puts forward that there were 13,319 number of vulnerabilities detected in 2019, in 1,607 apps.

Even though the security issues paint a very formidable problem, most of them can just as easily be combated by the knowledge of very key points. In this paper we will try to summarise the major issues faced and try to effectively solve them. In order to solve the problems practically and not just in theory, we have also created an E-Commerce application to imbue with all the security features we mention.

## 2. APPLICATION SECURITY ANALYSIS -

According to a survey conducted by 'owasp.org', the top web applications security threats of 2021 included Broken Access Control, Cryptographic failures and Injections among many others. All of these threats can be broken down into their potential target categories, namely client, server and data transmission.

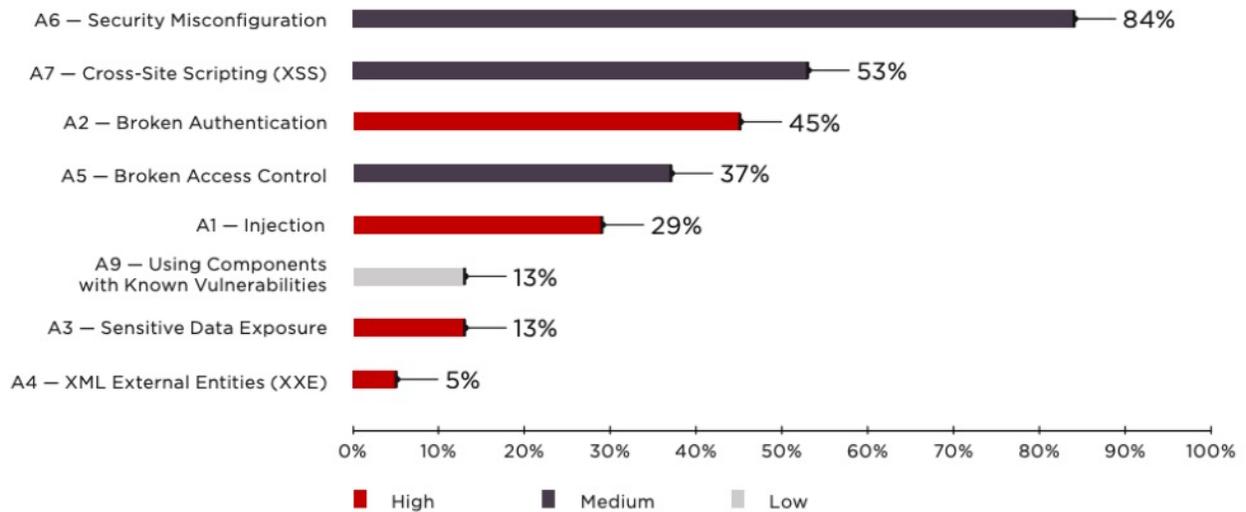

Fig 1. Most common OWASP Top 10 vulnerabilities 2019 (percentage of web applications)

## 2.1. CLIENT SIDE THREATS -

Client-side security, client-side vulnerabilities, and client-side attacks are terms used in cybersecurity to describe security incidents and breaches that occur on the customer's (or users') computer system rather than the company's (on the server side) or anywhere in between.These are plenty of ways in which the client side can be used to target users, some of which include -

### 2.1.1. CROSS-SITE SCRIPTING -

From the start, browsers were only meant to display HTML content but as the time went on developers were given more tools to make their web applications stand out. It began with the addition of Cascading style sheets(CSS) and eventually landed on Javascript. Javascript's purpose was to govern the functionality of the application. Naturally, because of its purpose it was immensely important and powerful. But, with this powerful way of increasing an application's functionality, it also meant opening a doorway for potential attacks. By inserting code, usually a client-side script like JavaScript, into a web application's output, cross-site scripting (XSS) attacks the application's users. The goal of XSS is to get client-side scripts in a web application to execute in the attacker's preferred way. XSS allows attackers to run scripts in the victim's browser, allowing them to hijack user sessions, deface websites, or redirect users to malicious websites.

### 2.1.2. MAGECART ATTACK -

Payment gateways have always been on top of the security priority list. Yet, client side attacks can still manage to steal extremely valuable information if the user is not careful. The Magecart attack is such an example.It's also known as payment skimming or e-skimming, and it works by inserting a small piece of code into a website and skimming information from clients. It's usually put into payment pages, where it functions as an online credit-card skimmer, collecting personal data and credit card information from everyone who comes across it. Attacks by magecarts have grown all-too-common. The following are some of the most well-known companies that have been impacted:British Airways, Ticketmaster, Newegg, Forbes and Macy's.

### 2.1.3. BROKEN LINK HIJACKING -

Broken Link Hijacking (BLH) is a web-based attack in which attackers take control of outdated, stale, and invalid external links on legitimate websites and web applications for malicious or fraudulent reasons. External links are utilised for a variety of reasons, including SEO and loading resources from external URLs/ sites. These links may expire or become invalid as a result of domain expiration, for example. And the attacker somehow gains control of the resource at the other end of the link and utilises it to further his or her goals.

## 2.2. SERVER SIDE THREATS -

An attacker (the client) launches a server-side attack directly at a listening service. Server-side attacks are attempts to hack and breach the data and applications on a server.

### 2.2.1. DENIAL OF SERVICE (DoS) -

Denial of Service, or DoS, is a server-side attack that effectively shuts down any website.. They accomplish this by increasing traffic on your site so much that the victim's server becomes unresponsive. While some DoS attacks come from single attackers, others are coordinated and are called Distributed Denial of Service (DDoS) attacks. Oftentimes, the users of computers executing a DDoS do not even know their computers are being used as agents. The number of DDoS attacks per day increased dramatically in the first half of 2021. In the first half of 2021, the average daily number of attack mitigations grew by 25% compared to Q4 of 2020. Mitigating an average of 1,392 attacks every day, with the highest number of attacks being 2,043 on May 24, 2021. During the first half of 2021, we mitigated a total of 251,944 distinct threats against global infrastructure.

### 2.2.2. BOTNET -

Attackers use botnets to automatically run and distribute malicious software on "agent" servers. They then use the agent machines to attack or infect others. Because all of this can be done automatically without user intervention, botnets can spread very quickly and be deadly for large networks. They are commonly used in DDoS attacks and spam campaigns.

### 2.2.3. UNPATCHED SOFTWARE -

Sometimes the biggest threat to a server can come from its own software. When the server's software starts ageing and isn't updated with the latest firmware, it can be an invitation for attackers to exploit the very exposed vulnerabilities. Rather obviously the solution to this is very simple, keep the server's software fully updated in order to leave any vulnerabilities exposed.

## 2.3. DATA TRANSMISSION THREATS -

The time between the client using the frontend to send its information to the point the server receives the data is known as the data transmission phase. This phase is especially susceptible to attacks as there is nothing directly governing this phase. Active attacks and passive attacks are the two sorts of attacks that can occur in this area. Active attacks mostly refer to network-based attacks. They mainly target to revise the entered data. Passive attacks are known to read the data in the network. They have access to the user's classified personal information and can extract it.

## 3. PROPOSED SOLUTION TO COMBAT THREATS -

## 3.1. CLIENT SIDE -

### 3.1.1. SQL INJECTION ATTACK -

The common use of SQL injection attack is to abuse web pages that allow users to input data into form fields for database queries. Injection is an unintended command sent to an interpreter. Attackers can enter the modified SQL query for user information. The queries directly communicate with the database for operations on data like data delete, create and change.

To protect our site from this we are passing user data to the server-side only after proper validation and authentication from the client-side.

### 3.1.2. URL INJECTION ATTACK -

A well-crafted attack url is often referred to as a query url. If we have a URL for a web page.For example: if you get an URL like

http://abc.&.in/test /pqr/pqr.html

That means we don't have any vulnerable sections on the page. But if the URL is like http://abc.s&.com/user.php?user_id='xxx' then the

query 'user id=xxx' is a string type query for the url, which an attacker can change.

To protect our site against the URL injection attack we are using Auth token so that to access any sensitive information one should have both the user id and the bearer token or the auth token.

### 3.1.3. REFLECTED XSS -

It occurs when an application receives data in an HTTP request and incorporates that data in an unsafe manner inside the immediate response.

To protect our site from this attack we are not using data directly from the URL bar or from the client side for the database queries. This in turn not only protects the user's data but also maintains the integrity of the developer's web application.

### 3.1.4. PASSWORD ATTACK -

Passwords are the most common technique of gaining access to a secure information system, which makes them a tempting target for cyber attackers. By accessing a person's password, an attacker can gain entry to confidential or critical data and systems, including the ability to manipulate and control said data/systems. Social engineering, gaining access to a password database, probing the network connection to obtain unencrypted passwords, or just guessing are all tactics used by password attackers to determine an individual password.

To protect our site against this we are storing the user password after hashing it in which we have applied 10 levels of salting and hashing.

### 3.2. SERVER SIDE -

### 3.2.1. BROKEN AUTHENTICATION -

When making any application, securing the user's personal data is the developer's first and foremost responsibility.Because of attackers' creativity, the system's poor design, and the inappropriate implementation of web apps, the chance of exploiting the Broken Authentication and Session Management vulnerability is rising. The consequence of the above exploitation may result not only in identity theft but also removal/tampering with confidential information. In our time trying to understand this issue, we came to the conclusion that it is most commonly caused due to the URL containing the session id and leaking it to someone else or the passwords not being encrypted either in storage or transit.

We tried to resolve both the issues by firstly not including session IDs in the URL but storing them in the local storage of users. The password encryption issue was dealt with fairly easily, we salted the password upto 10 times and tokenized it along the way. Even while storing the password exists in its token form. The password hashing and salting process is explained in Fig 2 and Fig 3.

### 3.2.2. SECURITY SETUP ISSUE -

Servers that are misconfigured from the beginning are more susceptible to attacks rather obviously. Here, misconfiguration refers to acts such as having directory listing enabled on the server, which might leak valuable information, running outdated software or not changing default keys and passwords. But the biggest setup issue that could arise could be revealing error handling information to the attackers. Every developer uses console logs in order to manage data flow in their application. Not removing these logs can lead to revealing error handling information and it is perhaps more common than it ought to be.

Just like with any other setup issue, its solution lies with updating the server apparatus as soon as a new stable update rolls out. Having a good 'build and deploy' process is also key. This process should be pre-defined to remove all server logs and any revealing information before going from development to production.

### 3.2.3. SENSITIVE DATA EXPOSURE -

Sensitive Data Exposure occurs when an organisation unknowingly exposes sensitive data or when a security incident leads to the accidental or unlawful destruction, loss, alteration, or unauthorised disclosure of, or access to sensitive data.While making backend APIs to get data from the database, it is extremely important that the APIs are well protected and only the people meant to access them can access them. A very common issue is that developers might forget to make their APIs exclusive. This can result in unauthorised people accessing these APIs which could quite tragically leak confidential user details.

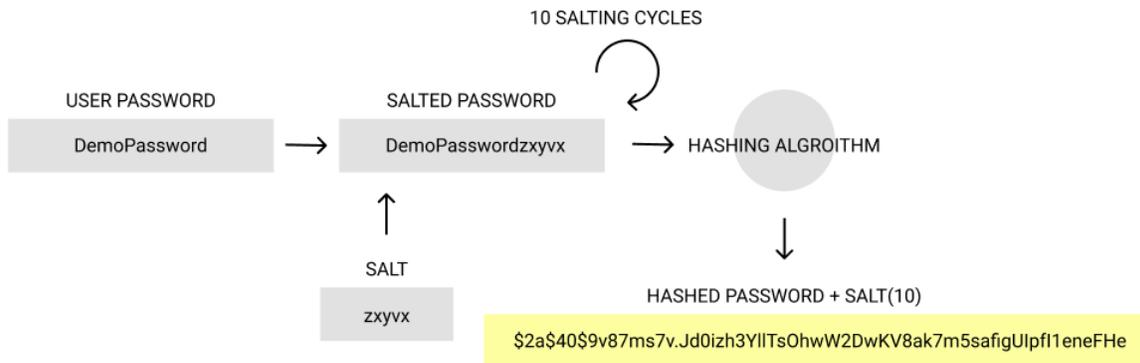

**Fig 2. Password Salting and Hashing (steps involved)**

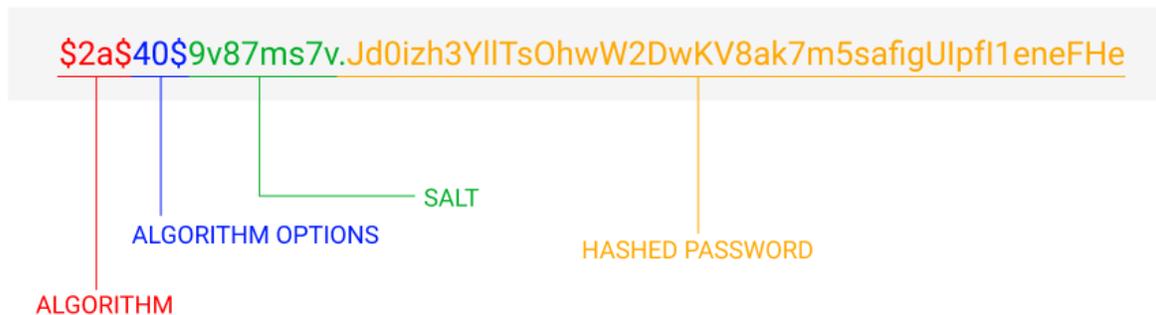

**Fig 3. After Processing Password Structure**

In real world use, we believe to have found a solution to help resolve this issue. In our practical application, we have added middlewares to APIs to limit their uses to only members of the application who should be logged in.

### 3.2.4. USING COMPONENTS WITH KNOWN VULNERABILITIES -

Just as the heading might suggest, being wary of issues arising from components with vulnerabilities. No developer is a stranger to NPM or PIP. They make development easy with reusable code written previously by developers from all around the world. But on closer inspection, they are just another Github repository with code to manage a specific task. There is a decent chance that the repositories that once were amazing to use might now be outdated and could possibly act as vulnerabilities in any application.

This is an issue without a very technical solution. To resolve this issue, the only thing to do is be careful. Always have a good look at the code you are planning to use before importing it in your own application.

Staying up to date with all the latest updates would also go a long way.

## 4. RESULTS -

Upon including all the preventive measures that were talked about to ensure a safe environment, the target web application should be ready. The talked about measures do not guarantee a completely safe application with zero vulnerabilities, but they do ensure that the application is safer than it was before all the aforementioned precautions were taken. The main threats discussed in the paper can be summarised to -

| THREAT | TARGET |
|---|---|
| CROSS-SITE SCRIPTING | CLIENT SIDE |
| MAGECART ATTACK | CLIENT SIDE |
| BROKEN LINK HIJACKING | CLIENT SIDE |
| DENIAL OF SERVICE | SERVER SIDE |
| BOTNET | SERVER SIDE |
| UNPATCHED SOFTWARE | SERVER SIDE |
| SQL INJECTION ATTACK | CLIENT SIDE |
| URL INJECTION ATTACK | CLIENT SIDE |
| REFLECTED XSS | CLIENT SIDE |
| PASSWORD ATTACK | CLIENT SIDE |
| BROKEN AUTHENTICATION | SERVER SIDE |
| SECURITY SETUP ISSUE | SERVER SIDE |
| SENSITIVE DATA EXPOSURE | SERVER SIDE |
| USING COMPONENTS WITH KNOWN VULNERABILITIES | SERVER SIDE |

**Table 1. THREATS CONCLUSION**

## 5. CONCLUSION -

This paper covers the problem faced in the modern world by users and developers alike and tries to come up with a solution to effectively resolve the threats that may arise.

We have discussed the top problems faced at the Client side and Server side along with the threats associated with data transmission as well. With the mentioned problems we have also proposed solutions that we have found in practice to help combat the issues.

## 6. CONCLUDING REMARKS -

With the internet playing such a crucial role in the lives of people nowadays, it has become important to analyse the threats posed to our applications on the said internet. At first glance it may seem the malicious threats outweigh the solutions proposed to resolve them. But, with passing time developers have come up with solutions to resolve the issues in the most effective ways possible. But, it remains true without being said that all the applications need to be monitored continuously and have to be evaluated at regular intervals to ensure proper safety of applications.

## 7. REFERENCES -